# Explanation for the development of surface -plasma method of negative ion production in BINP and around the World.

## Vadim Dudnikov

Development of charge-exchange ions injection into circular accelerators and storage rings and development of surface-plasma method of negative ion production in Budker Institute of Nuclear Physics (Novosibirsk) and around the World are described.

In the first and second years at Novosibirsk State University, physics seminars in our group were conducted by Gennady Ivanovich Dimov. After the second year, in 1962, he invited me to the laboratory of the Institute of Nuclear Physics to work on the development of charge-exchange injection of protons into ring accelerators.

G. Budker proposed charge-exchange injection to obtain high-brightness beams for the VAPP-4 proton-antiproton colliding beam project (this program was later implemented at Fermi National Accelerator Laboratory, USA, using developments from the Institute of Nuclear Physics (INP) in charge-exchange injection, antiproton production, and electron cooling, as well as with significant participation from former members of BINP [ ]). Since the goal was to accumulate beams with maximum intensity, and the record intensity of $H^-$ ion beams at the time was only 70 μA, an important part of the program was the development of methods for obtaining intense beams of negative hydrogen ions (several mA).

Initially, I worked on measuring the stripping cross-sections of $H^-$ and $H^0$ in gas targets to achieve maximum $H^0$ yield at energies up to 1.5 MeV. This became my diploma thesis. Today, these data are used to optimize the yield of neutrals from accelerated negative ions in neutral beam injectors for fusion devices [G.I. Dimov and V.G. Dudnikov: Zhur. Tekh. Fiz, 36, (1966) 1239; Sov. Phys. -Tech. Phys. 11, 919 (1967)]. After graduating from NSU in 1965, I entered the graduate program at the Institute of Nuclear Physics and began working on an experiment to accumulate an intense proton beam through charge-exchange injection in a small storage ring. The setup scheme for studying charge-exchange injection is shown in Fig. 1.

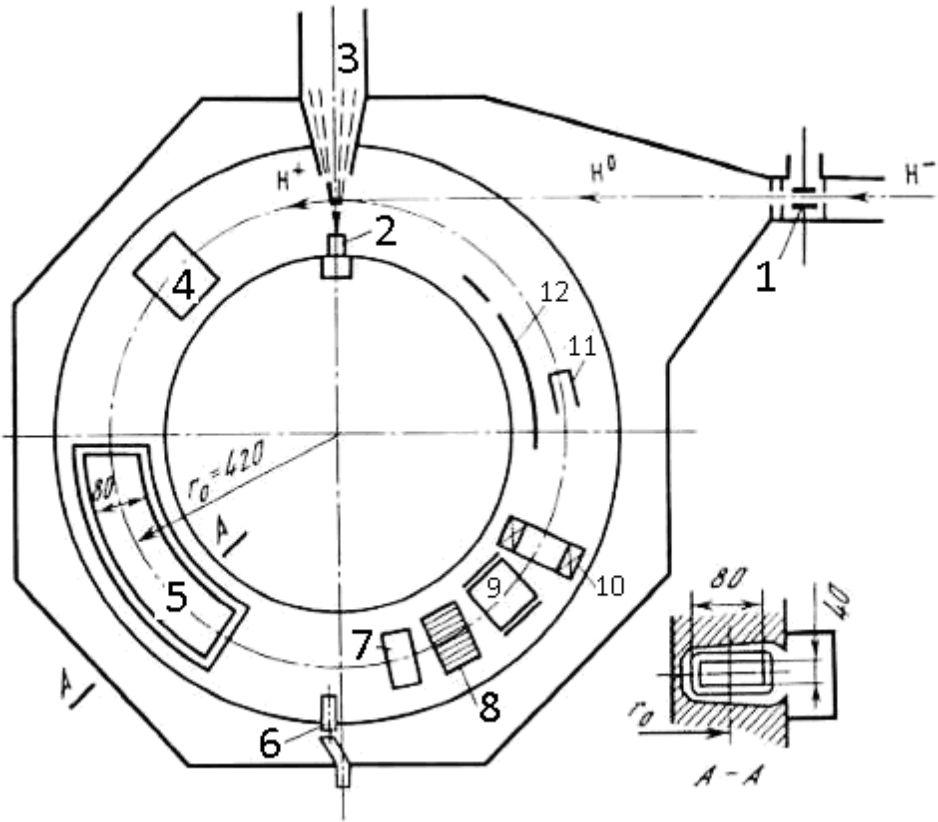

**Fig. 1. Diagram of the setup for studying charge-exchange injection**
1 - first stripping target, 2 - supersonic jet nozzle, 3 - jet receiver, 4 - ring pickup electrode, 5 - RF acceleration drift tube, 6 - collimator of the fluorescent beam profiler, 7 - ionization beam intensity meter, 8 - ionization profile monitor, 9 - beam position pickup, 10 - Rogowski belt, 11 - Faraday cylinder, 12 - deflector for suppressing electron-proton instability.

At the first stage of work, it was necessary to launch the electrostatic accelerator at an energy of up to 1.5 MeV with an arc H$^-$ ion source (of the Ehlers type) with a current of up to 1 mA, which was quite large for that time. There were significant problems with starting the system, but through persistent effort, these were overcome, allowing the accumulation of protons with the injection of +0.5 mA of neutrals into orbit. Initially, the intensity of the accumulated beams was limited by the excitation of coherent synchrotron oscillations due to ionization losses in the hydrogen stripping jet, and later by the negative mass effect and the application of RF voltage. After overcoming these effects, it became possible to accumulate up to 300 mA of circulating protons, limited by longitudinal space charge.

A high-intensity beam remained stable for 1-5 milliseconds after accumulation, but then radial betatron oscillations rapidly developed, causing the beam to be lost to the walls within a few revolutions. The time before the onset of instability depended on the beam intensity, target thickness, gas density in the chamber, RF voltage amplitudes, and voltages on the electrodes surrounding the beam. This instability was successfully suppressed using feedback.

Various instability models known at the time were considered to explain this unusual instability. However, no known instability matched the observed behavior.

At the same time, in a neighboring research section, under the leadership of B. Chirikov, work was being carried out on the accumulation of an intense electron beam in the Betatron B-3 with spiral accumulation to produce a stabilized electron beam, a concept proposed by Budker back in 1953. They managed to accumulate up to 300 A of electrons, but during acceleration, the beam was lost to the walls. To explain these effects, Chirikov developed a theory of the stability of an electron beam partially compensated by ions. The ions oscillate in the potential of the electron space charge, creating electric fields that drive transverse oscillations of the circulating electrons. The oscillations of the circulating electrons, in turn, create electric fields that excite transverse oscillations of the ions. If the frequencies of these oscillations coincide, the oscillations grow rapidly, converting a large portion of the longitudinal motion energy into transverse oscillation energy.

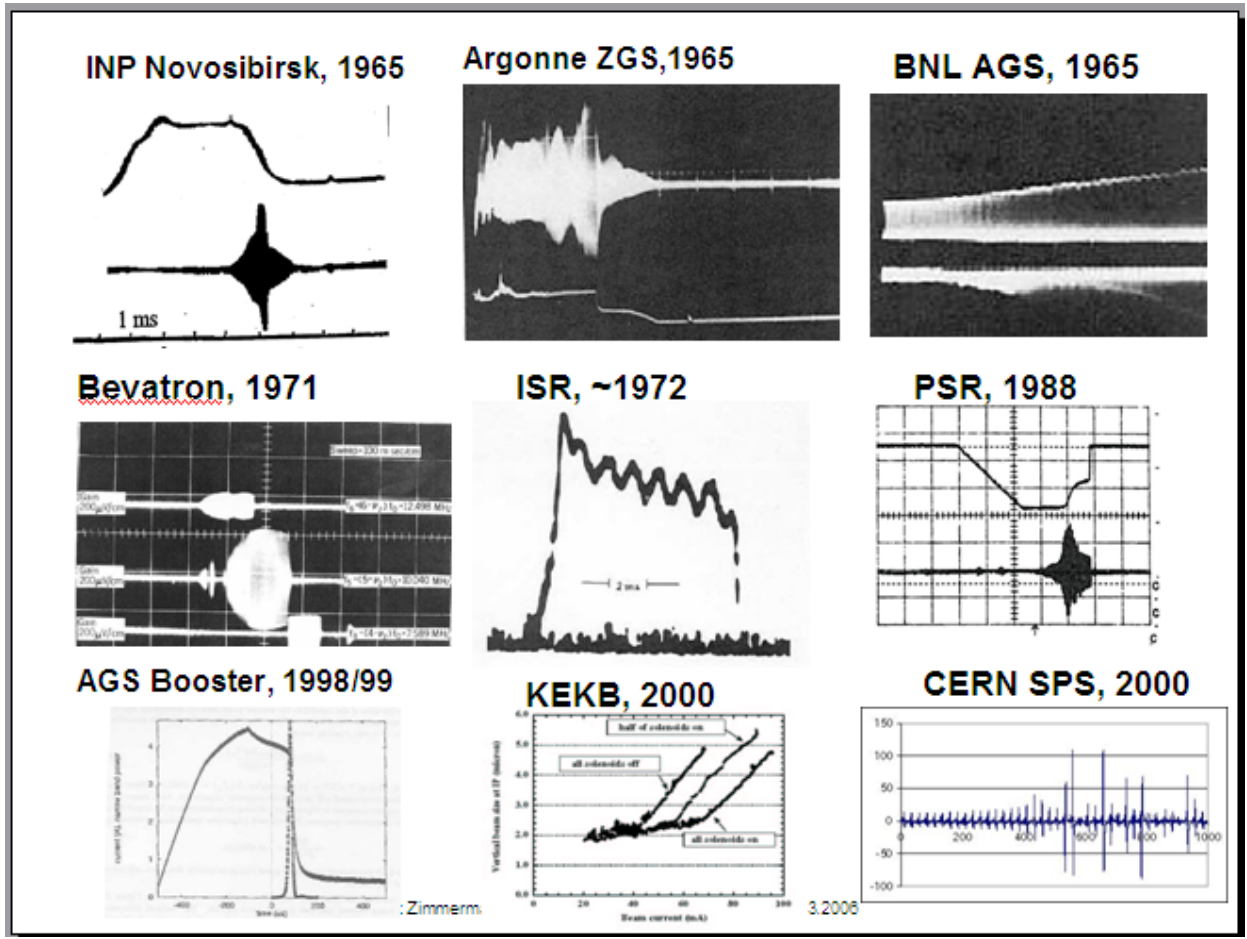

**Figure 2. History of the Discovery of e-p Instability Worldwide**

Chirikov lectured on physics in our course, and since we worked nearby, I told him about our instability and my hypothesis that it might be related to proton beam compensation by electrons. The question of whether a heavy proton beam could be destabilized by light electrons was resolved positively. For low-mode oscillations, the amplitude of electron oscillations is much greater than that of proton oscillations, allowing electrons to be ejected from the beam. However, at higher frequencies, the amplitudes become comparable, and the instability can grow until the protons are lost.

Later, a more detailed theory of this e-p instability was developed by Koshkarev and Zenkevich.

The candidate dissertation **"Production of an Intense Proton Beam in a Storage Ring Using Charge-Exchange Injection"** (supervised by G.I. Dimov) was prepared in 1966 and defended in 1967. The opponents were A. Skrinsky, Yu. Ado, and A. Naumov.

This e-p instability, later named the **electron-cloud effect (ECE)**, was rediscovered multiple times. It was revisited 30 years later when electron-positron and proton colliders with a large number of bunches, such as **B-factories, LANSCE, SNS, and LHC**, began operating.

*Fighting of electron cloud effect is an important component in the development, commissioning and upgrading of all modern accelerator projects. "Although the electron-cloud effect (ECE) for lepton beams was first reported in 1995 [1], the ECE was apparently first discovered 30 years earlier at the Budker Institute (Novosibisrk) as a form of a two-stream instability for coasting proton bunches [**V. Dudnikov**, "The intense proton beam accumulation in storage ring by charge-exchange injection method," Ph.D. thesis, Novosibirsk INP,1966] " (Miguel Furman, 2004, **mafurman.lbl.gov/ECLOUD04_proceedings/furman-goals-LBNL-57166.pdf** ).*

«The first observations of electron-proton coupling effect for coasting beams and for long-bunch beams were made at the earliest proton storage rings at the Budker Institute of Nuclear Physics (BINP) in the mid-60's [1]. The effect was mainly a form of the two-stream instability. This phenomenon reappeared at the CERN ISR in the early 70's,where it was accompanied by an intense vacuum pressure rise. When the ISR was operated in bunched-beam mode while testing aluminum vacuum chambers, a resonant effect was observed in which the electron traversal time across the chamber was comparable to the bunch spacing [2]. This effect ("beam-induced multipacting"), being resonant in nature, is a dramatic manifestation of an electron cloud sharing the vacuum chamber with a positively-charged beam. An electron-cloud-induced instability has been observed since the mid-80's at the PSR (LANL) [3]; in this case, there is a strong transverse instability accompanied by fast beam losses when the beam current exceeds a certain threshold. The effect was observed for the first time for a positron beam in the early 90's at the Photon Factory (PF) at KEK, where the most prominent manifestation was a coupled-bunch instability that was absent when the machine was operated with an electron beam under otherwise identical conditions [4]. Since then, with the advent of ever more intense positron and hadron beams, and the development and deployment of specialized electron detectors [5-9], the effect has been observed directly or indirectly, and sometimes studied systematically, at most lepton and hadron machines when operated with sufficiently intense beams. The effect is expected in various forms and to various degrees in accelerators under design or construction. The electron-cloud effect (ECE) has been the subject of various meetings [10-15]. Two excellent reviews, covering the phenomenology, measurements, simulations and historical development, have been recently given by Frank Zimmermann [16,17]. In this article we focus on the mechanisms of electron-cloud buildup and dissipation for hadronic beams, particularly those with very long, intense, bunches.» Фурман, LBL.

After accumulating intense beams in synchrotron mode, the accumulation of continuous beams was initiated, with compensation for ionization losses using an induction field to obtain beams with space charge compensation. Eventually, the instabilities were suppressed, and proton beams with an intensity of 1 A were accumulated—9 times higher than the space charge limit and 150 times above the threshold for e-p instability growth. These results are described in the second edition of Martin Reiser's book *"Theory and Design of Charged Particle Beams"* (2006).

*"This was one of the most remarkable discoveries made at the Institute of Nuclear Physics. Thank you for the article,"* wrote D.D. Ryutov in 2019.

In 1968, Ronald Martin, then head of the Zero Gradient Synchrotron (ZGS) at Argonne, visited the Institute of Nuclear Physics. After becoming familiar with charge-exchange injection, he concluded that it would help him compete in intensity with the AGS at BNL and began implementing charge-exchange injection at Argonne. This story is vividly described by him and his team in «History of the ZGS 500 MeV Booster, ANL-HEP-TR-06-446*April 2006,*by J. Simpson, R. Martin, and R. Kustom High Energy Physics Division, Argonne National Laboratory, April 2006, http://www.ipd.anl.gov/anlpubs/2006/05/56304.pdf.

Charge-exchange injection was successfully implemented at ZGS and its booster, which for many years remained an intense pulsed neutron source with record-breaking parameters. After that, charge-exchange injection was adopted at FNAL, BNL, KEK, LANL, RAL ISIS, DESY HERA, CERN booster, and SNS ORNL.

Charge-exchange injection is used in the CELSIUS storage ring (Uppsala, Sweden) [ ], in the COSY storage ring (Jülich Research Center, Germany). It was also utilized in the synchrotron of the Institute for Theoretical and Experimental Physics (ITEP) for the accumulation of carbon ions.

Currently, work is underway to transition to charge-exchange injection at U-70 in Protvino and in the CERN Booster. Stripping of negative ions is also used for extracting accelerated ions from thousands of cyclotrons for the production of medical isotopes.

Yura Belchenko, as a student, was passed down to me from Yura Kononenko. They were working on improving the proton source with a diaphragm arc discharge and a grid beam formation system. This source had been developed by G. Dimov and was used to generate H⁻ ions via the charge-exchange method.

After Y. Kononenko left for Kyiv, G. Dimov temporarily assigned me to supervise this work. At that time, a new lab technician, Gena Markov, was hired. He was highly skilled with his hands, and together with P.A. Zhuravlev, they engaged in inventive work.

In Dimov's arc source, ions were accelerated between two grids made of thin wires with a diameter of 0.1 mm, spaced about 0.5 mm apart, and with a gap of ~2 mm, precisely aligned with each other. With an ion current density of ~1 A/cm², careful conditioning allowed the accelerating voltage to be increased up to 15 kV. However, during breakdowns, the cathode wires would often burn out, making practical use of this high-current-density source, with its record-low ion temperature (~0.005 eV), impossible.

At my suggestion, Gena Markov took on the task of creating a precision cathode grid by replacing the wires with thin strips. The first louvered grids were made using a complex technology that involved spark-cutting a briquette of welded louvers in a frame. These grids demonstrated high resistance to breakdowns, making Dimov's source practically viable. Later, P.A. Zhuravlev developed a more cost-effective method for manufacturing louvered grids, and Grisha Roslyakov continued working with these sources.

Based on this technology, the first diagnostic neutral beam injectors for plasma installations were developed. Over time, more powerful injectors were designed for many plasma installations worldwide. These injectors helped the plasma researchers at the Institute of Nuclear Physics (INP) survive difficult times and carve out a niche in global scientific and technological collaboration, as well as in international competition.

At that time (1970), M.A. Lavrentyev was actively opposing Budker's policy of expanding the Institute of Nuclear Physics (INP), and Budker was searching for ways to ensure the institute's survival. After the experimental validation of charge-exchange injection (which became the basis of my Ph.D. dissertation and G.I. Dimov's doctoral dissertation), Budker proposed using high-energy neutral beams in space to impact cosmic objects—such as inspecting satellites for nuclear materials and suppressing nuclear explosions.

At the Korolev Central Design Bureau (which was already operating without Korolev), funds left unused after the closure of the Soviet lunar program were redirected, and an agreement was signed with INP to develop a neutral beam injector capable of producing an H⁻ ion current of 10 mA, with a budget of 2 million rubles (at that time, a nine-story apartment building with 2,048 apartments was cost 1 million rubles, 1$ was 086 rubl).

Three research groups were formed to explore different methods of generating negative ions. In Dimov's laboratory, charge-exchange and plasma-based negative ion sources (similar to the Ellers type) were being developed for charge-exchange injections.

In my first year of university, I read Gaponov's textbook *Electronics* and later revisited it. In it, I found a mention of secondary emission of negative ions—an effect that was little known at the time.

I decided to study secondary ion-ion emission in more detail. Although there was a considerable amount of data, it was highly contradictory, and everywhere the probabilities of secondary emission were very low. The theoretical foundations of this phenomenon were also absent. It was known that depositing alkali metals increased the coefficient of secondary ion-electron emission. I came across works by Ayukhanov (1961) and Kron (1962), which showed that when alkali metals were deposited, the coefficient of secondary negative ion emission also increased—but the resulting $H^-$ ion currents were at a sub-microampere level.

I suggested to Dimov that we should investigate the secondary-emission method of producing negative ions. He enthusiastically supported the idea, and a group was formed under his leadership to generate $H^-$ ions by bombarding surfaces with cesium ions. I actively participated in this work. A test stand was built, and soon we were able to generate $H^-$ beams with currents up to -2.5 mA. However, the brightness of these beams was very low, and the lifespan of these devices was extremely short.

Meanwhile, G. Roslyakov's group focused on charge-exchange sources, while Y. Belchenko and I worked on plasma sources. All the groups were working intensively, but we were still far from achieving the required parameters.

Budker held weekly meetings where various ideas were discussed, but no viable solutions were found. By the end of the contract, everyone had lost hope of achieving the necessary ion currents and left for summer vacation. Yura went with a construction team to Bilibino in Kolyma to help build the Bilibino Nuclear Power Plant, taking our lab technician, Dima Melnichuk, with him. The mechanics also went on vacation.

I stayed alone over the summer and continued working with the plasma source, which had a planotron geometry. It was capable of producing up to 4.5 mA of $H^-$ ions in 1-millisecond pulses, but this required an electron current to the extractor that was 50 times higher, a discharge current of 100 A, and a discharge voltage of 600 V. The discharge power of 60 kW was inserted to a volume ~1 $cm^3$.

On July 1, 1971, I attached a cesium chromate pellet containing 1 mg of cesium to the anode of the planotron and turned on the discharge. The emission slit was shielded, and the collector registered an ion current of 1.5 mA. After a few minutes of operation, a current spike appeared at the end of the pulse, reaching up to 3 mA.

After optimizing the gas flow, the current at the end of the pulse increased to 4 mA. However, after 20 minutes, the current spike disappeared, and the collector current dropped back to 1 mA.

Believing that the current spike was related to cesium release, I placed several pellets on the cathode, which was heating up more intensely, and covered them with a nickel mesh.

In this configuration, the collector current quickly increased to 12 mA. After further optimization of the gas flow and discharge, I achieved a stable rectangular pulse of 15 mA.

The discharge voltage dropped from 600 V to 100 V.

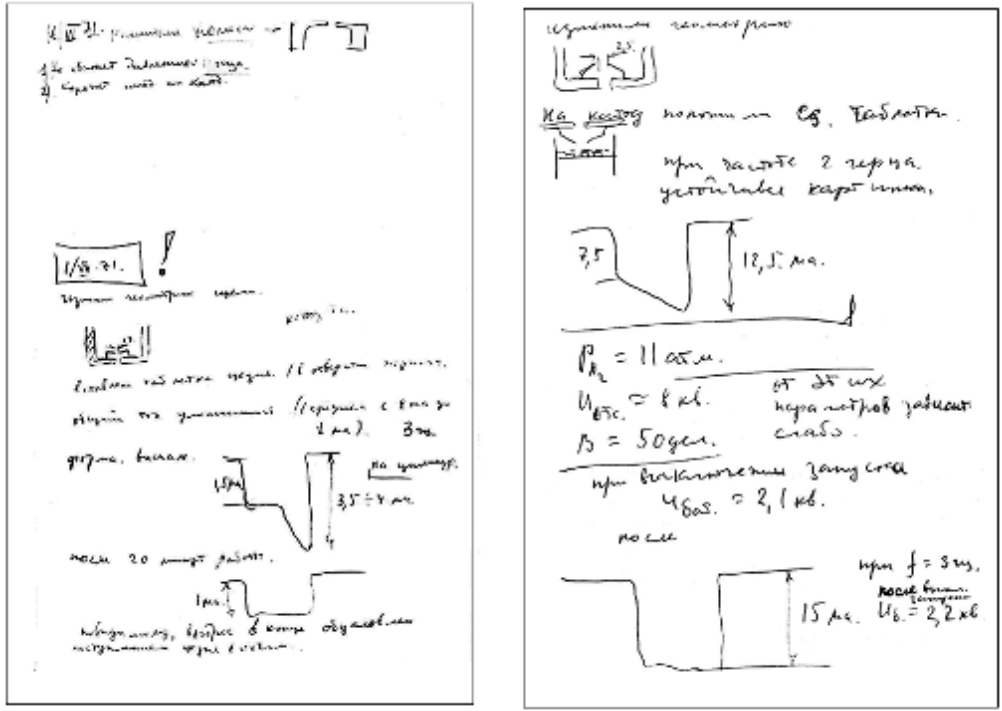

**Fig. 3. A copy of a page from the work journal dated July 1, 1971, describing the evolution of the negative ion beam intensity when cesium was added to the discharge.**

Figure 3 presents an entry in the work notebook about this event.

After this, various discharge configurations were tested over the course of a week, confirming that the main particles were H⁻ ions, while electron and heavy ion currents on the collector were minimal. After ensuring everything was in working order, I left for vacation to my village, Gunda, in Buryatia.

On the way back, I developed appendicitis on the bus, but the surgery was performed in time, and I had to spend another two weeks recovering.

Upon my return, Y. Bulchenko and G. I. Dimov attempted to restart the source with cesium, but they were unable to do so before I arrived. It is worth noting that, later on, almost no one could independently achieve the enhanced negative ion generation with cesium—only after being shown the process in a step-by-step demonstration. Y. Bulchenko later trained both French and Japanese researchers in the cesiation technique.

Once I returned, the source functioned properly, and the H⁻ current quickly increased first to 100 mA, then to 300 mA, and finally up to 0.9 A from a source no larger than a cigarette lighter.

Through an agreement with the Central Design Bureau (ЦКБ), we successfully reported our progress and secured new contracts worth 5 million rubles each—one for the production of SPS for the linear accelerator of the meson factory at the Institute for Nuclear Research (IYI) and another for the Institute of High Energy Physics (ИВФЭ). (For context, at that time, a nine-story apartment building with 248 apartment cost 1 million rubles.)

At that point, the cesiation results were classified as secret, and publication was prohibited. However, many delegations from across the Soviet Union (ЦКБ, НИИЭФА, ИНР, ИФВЭ) and from the United States began visiting ИЯФ. Budker allowed high-ranking U.S. visitors to view the SPS as potential trade objects.

During this period, I spent a lot of time working at the ЦКБ in Podlipki with Mikhail Vasilyevich Melnikov's group and Ivan Iosifovich Rayko. They shared many stories about working with Korolev.

Before our publications, articles began appearing stating that negative ion yields could be increased by adding cesium to the discharge:

**IEEE TRANSACTIONS ON NUCLEAR SCIENCE**
**Volume: NS20, Issue: 3, Pages: 136-141, DOI: 10.1109/TNS.1973.4327065, Published: 1973**
*"SOME ADVANCES IN NEGATIVE ION TECHNOLOGY" by Kenneth H. Purser*

*"During the past two decades, the advances which have been made in the techniques of negative ion production have been truly spectacular, both in terms of intensity and the available species. For example, while in 1955 negative ions of hydrogen isotopes were produced at sub-microampere intensities, today 22 mA beams of negative hydrogen have been reported [2], and it is rumored that Dimov and his collaborators at Novosibirsk have seen 200 milliampere peak pulse intensity during pulsed H⁻ operations."*

In reality, by that time, we had already achieved an H⁻ ion beam current of 0.9 A.

In the final version of the review by K. Prelec and Th. Sluyters, *"Formation of Negative Hydrogen Ion in Direct Extraction Sources"* (Rev. Sci. Instrum, 44 (10), 1451, 1973), the following was stated without any explanations:

*(These conclusions were absent in the preprint and were added at the last moment before publication, based on rumors or other sources of information.)*

In the final version of the review by **K. Prelec and Th. Sluyters**, *"Formation of Negative Hydrogen Ions in Direct Extraction Sources,"* **Rev. Sci. Instrum, 44 (10), 1451, 1973**, the following statement was made without any explanations:

*(These conclusions were absent in the preprint and were added at the last moment before publication, based on rumors or other information).*

K. Prelec and Th. Sluyters did not claim the discovery of cesiation and later referenced our preprints. However, these mentions of cesiation remain among the earliest available publications, apart from my **Author's Certificate No. 411542**.

After confirming the significant effect of cesiation, **Dimov advised me to patent the process**, because, as he said at the time, *"In about 20 years, some guy will come along and claim he did it all himself."*

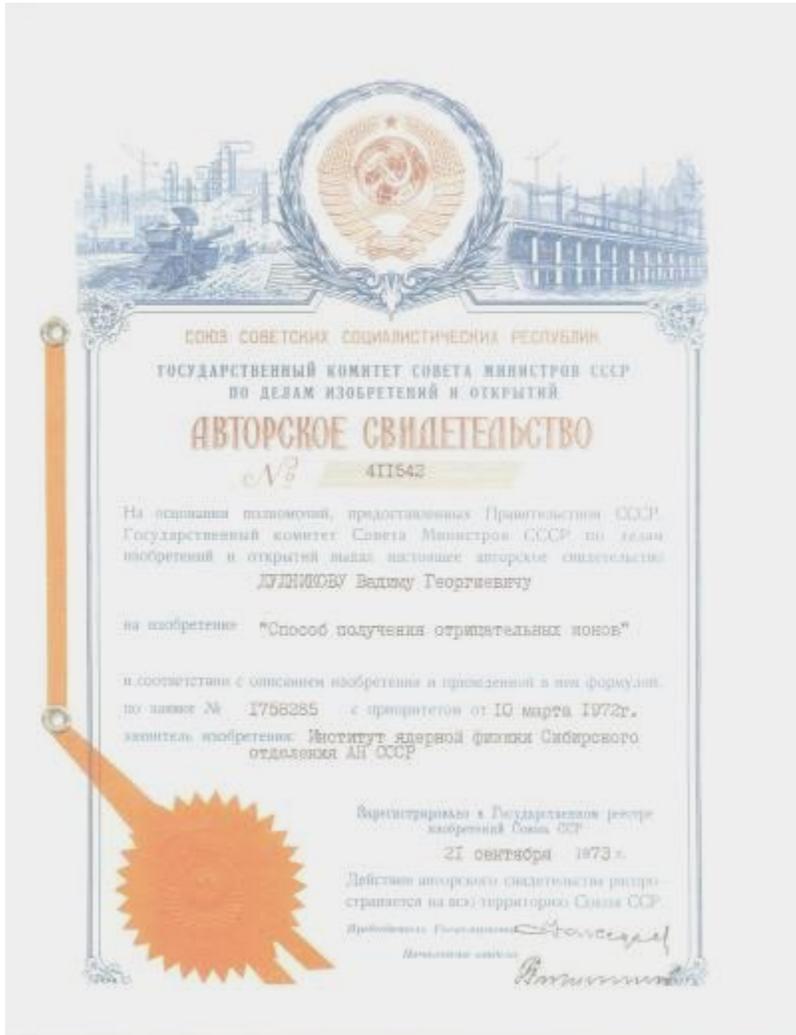

Figure 4. Author's Certificate No. 411542

Dimov was right. Many were excited upon witnessing the effect of cesiation and tried to present themselves as the pioneers without referencing our priority works. Even now, most citations refer to secondary studies that replicate our results but fail to acknowledge them.

Gennady Ivanovich always emphasized in conversations that the discovery of the cesiation effect was made personally by me. As a result, old visitors remember Dimov's words:

**From Ronald Martin, ZGS, Argonne NL:** *'Dimov was a leader of the ion source group at Novosibirsk, which included another physicist by the name of Dudnikov. Dudnikov has become well known worldwide by extending the $H^-$ ion source development to exceed 100 mA.'*

At that time, an **Author's Certificate** was issued:
V. Dudnikov, "Method for Producing Negative Ions," *Author's Certificate No. 411542, C1.H013/04, filed March 10, 1972.*
Link to INIS database**"A method for producing negative ions from a gas-discharge source, characterized by the fact that, in order to increase the efficiency of the source, a substance with a low ionization potential, such as cesium, is introduced into the discharge chamber along with the primary working substance."**

Eventually, permission was granted to publish in *ZhTF* (Journal of Technical Physics) regarding a **planotron** with a 22 mA current *without* cesium, even though at that time, sources with cesium were already achieving **1 A**. It was only in 1974 that authorization was given for an international publication:

**Y. I. Belchenko, G. I. Dimov, and V. G. Dudnikov,** *"A powerful injector of neutrals with a surface-plasma source of negative ions,"* **Nucl. Fusion 14(1), 113 (1974)** [Available now].

Based on this cesiation effect, together with **G. I. Dimov and Y. I. Belchenko**, we developed the **fundamental principles of efficient negative ion generation**. It was demonstrated that cesiation **greatly enhances the secondary emission** of negative ions from surfaces bombarded by plasma particles. The **reduction in surface work function** due to cesiation significantly increases the probability of sputtered and reflected particles escaping as negative ions.

The **theory of negative ion emission dependence on work function and particle energy** was developed by **M. Kishinevsky**. This led to the discovery and development of an entirely new method—**the surface-plasma method (SPM) for efficient negative ion generation**—and the creation of **surface-plasma sources (SPS)** with cesiation, ensuring highly efficient negative ion production.

Key publications on this work include:

- **Y. I. Belchenko, G. I. Dimov, and V. G. Dudnikov,** *"A powerful injector of neutrals with a surface-plasma source of negative ions,"* **Nucl. Fusion 14(1), 113 (1974)**

- *"Physical Principles of the Efficient Negative Ion Production,"* Symposium on the Production and Neutralization of Negative Hydrogen Ions and Beams, Brookhaven, 1977 (Brookhaven National Laboratory, BNL-50727, pp. 79–96)
- **Y. Belchenko and V. Dudnikov,** *"Surface negative ion production in ion sources,"* in *Production and Application of Light Negative Ions,* 4th European Workshop, edited by W. Graham (Belfast University, 1991), pp. 47–66
- **V. Dudnikov,** Proc. *Second Symposium on Production and Neutralization of Negative Hydrogen Ions and Beams*, Brookhaven, 1980 (BNL-51304, p. 137)
- **V. Dudnikov,** *"20 years of surface plasma sources development,"* Rev. Sci. Instrum. **63, 2660 (1992)**
- **V. Dudnikov,** *"Thirty years of surface plasma sources for high-efficiency negative ion production,"* Rev. Sci. Instrum. **73, 992 (2002)**

Our results on cesiation were published **with significant delay and very limited distribution** because they were classified due to the **initiation of the "Star Wars" program**.

High-brightness **H⁻ beams for accelerators** were obtained from **SPS with noiseless discharges and electron oscillations in a magnetic field**—these later became known as **Dudnikov Sources**.

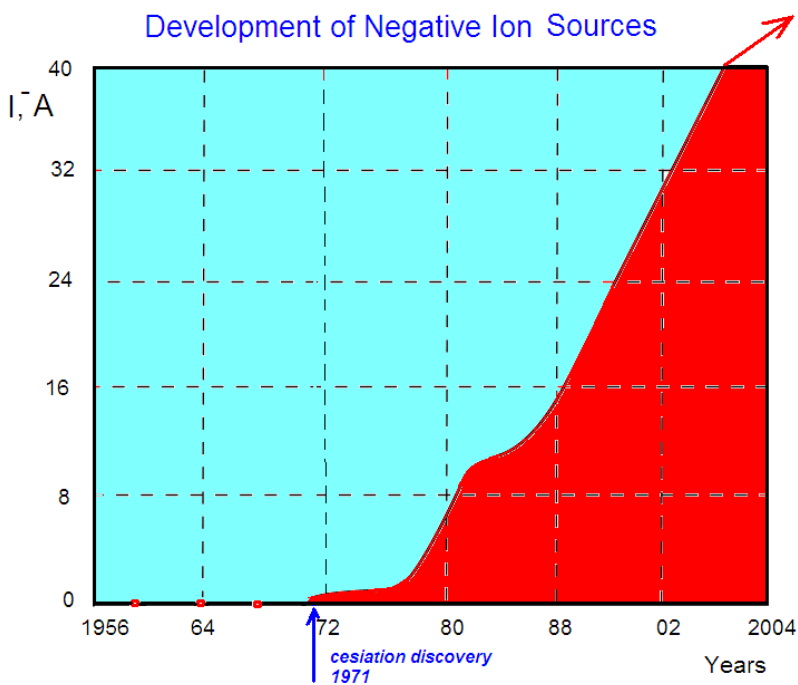

Fig. 5. **Intensity of Negative Ion Beams: 1971-discovery of Cesium Catalysis**

**V. Dudnikov,** *"Surface-Plasma Negative Ion Source with Penning Geometry,"* Proceedings of the **4th All-Union Conference on Charged Particle Accelerators**, Moscow, *Nauka*, Vol. 1, p. 323, 1974.

This paper has been widely cited, but it is not indexed in **Web of Science**.

The further development of **high-brightness SPS** was carried out in collaboration with **G. Derevyankin and V. Klyonov**. These studies and technological developments were rapidly adopted by **all U.S. National Laboratories**, as well as research institutions in **Europe and Japan**.

In the **U.S. and Japan**, hundreds of millions of dollars were allocated for the implementation of SPS technology. Within a short time, **the intensity of H⁻ ion beams increased by a factor of $10^4$**, from the previous record of **3 mA** to over **40 A**.

At that time, many researchers in the **U.S. and Japan** began learning Russian to read the latest **preprints and journal articles** in Russian, although these documents were quickly translated into English.

In **1972**, **Paul Allison** at **Los Alamos** began developing an **H⁻ ion source** for the **Los Alamos Meson Factory (LAMF)** to generate **short, high-power neutron pulses** via **charge-exchange injection of H⁻ into a compact proton accumulator**.

Initially, **Allison** attempted to replicate **Dimov's charge-exchange source**, but without success.

After the **publication of the SPS with Penning Discharge in 1976**, Allison learned Russian and **traveled with his wife to Novosibirsk** via Japan and **Vladivostok on the Trans-Siberian Railway**.

They lived in **Akademgorodok** for a **month**, during which **he was granted permission to study the work on SPS**.

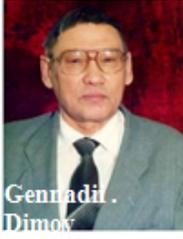

**Fig. 6. Slide from Martin Stockley's Presentation**

Upon returning to the **U.S.**, **Allison secured a $50 M grant** to **replicate the "Dudnikov Source."**

The development of **surface-plasma sources (SPS)** was carried out in **Paul Allison's group** for **12 years**, in parallel with similar efforts in **all other U.S. National Laboratories**. However, achieving **stable operation** of **Dudnikov source clones** in **low-noise regimes with high beam brightness** proved difficult.

At the **Los Alamos Neutron Science Center (LANSCE)**, they still use an **SPS with a converter**, where the **cesium consumption** is **~1 g/day**, whereas in **well-optimized SPS**, the cesium consumption is **~1 mg/day**.

In Los Alamos, they developed **Dudnikov's source** and **Tepljakov's RFQ accelerator** in a **flight-ready configuration**, testing them in the **BEAR experiment on a rocket**.

The **"Dudnikov Source"** was successfully adapted at the **Rutherford Appleton Laboratory** for the **ISIS neutron source**, where it has been operating **successfully for 30 years**.

All these results formed the basis of the **doctoral dissertation**:

📄 **V. Dudnikov,** *"Surface-Plasma Method for Generating Negative Ion Beams,"* **defended in 1978**.

Several **SPS with record-breaking performance** (**Dudnikov sources**) were supplied to **various institutes of the USSR Academy of Sciences and other organizations**.

Later, **high-brightness SPS** were developed for **international deliveries to U.S. universities and laboratories**.

Currently, at **Budker Institute of Nuclear Physics (BINP)**, a **high-power neutral beam injector** based on **SPS** is under development for **plasma injection** (**Yu. Belchenko**, **10A H⁻, 1 MeV**, Fig. 7).

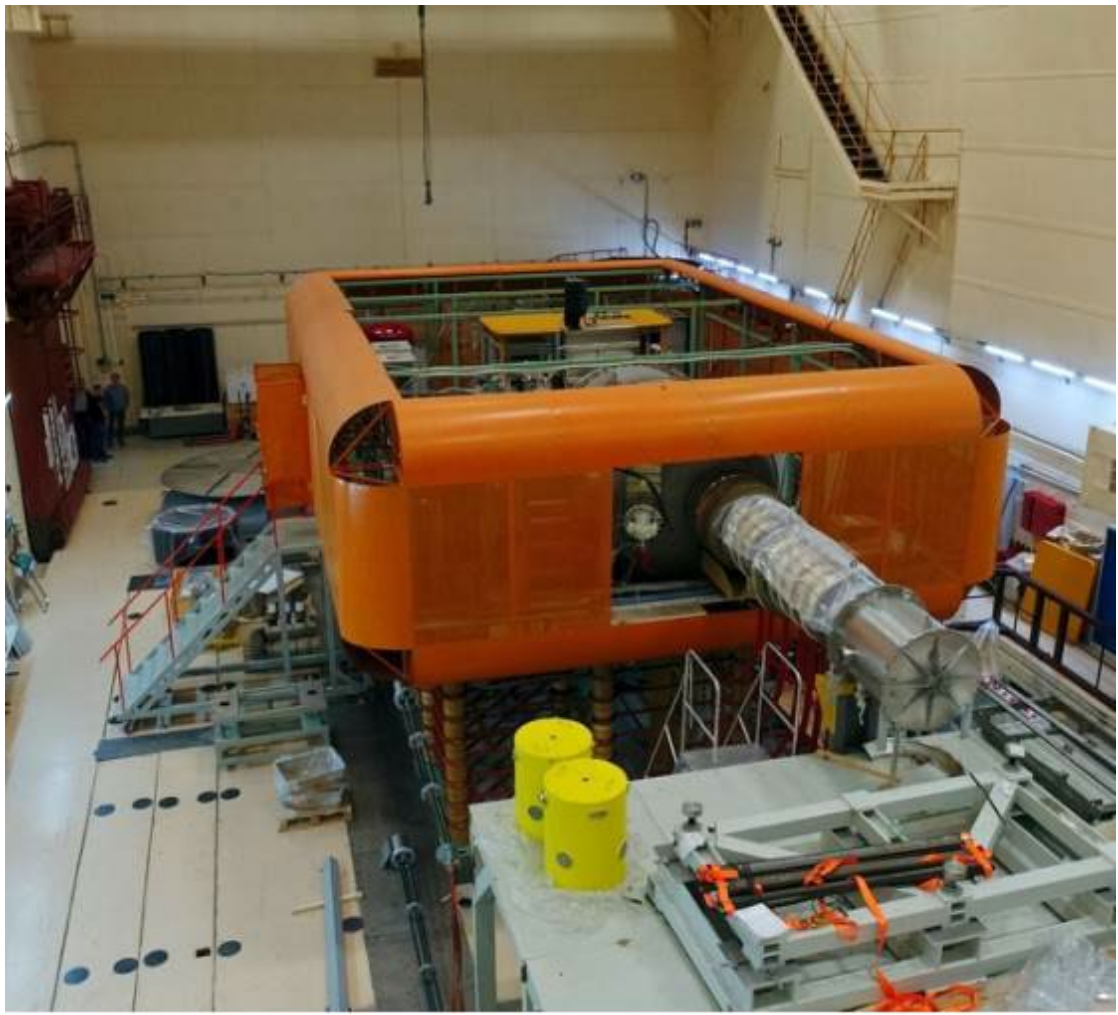

**Fig. 7. Photograph of the Neutral Beam Injector with RF SPS in BINP (9 A, 1 Mv)**

For a long time now, **publications on SPS in Russian have been almost nonexistent**, and there are **very few readers left** who can understand them.

Meanwhile, **large research coalitions** in the **U.S., Europe, Japan, India, Korea and China** continue active development of **SPS for accelerators and fusion energy applications (UFS)**.

Unfortunately, **older Russian-language publications are not available online**, but there is a **comprehensive review covering almost all major works**:

📖 **Nikita Wells,** *"The Development of High-Intensity Negative Ion Sources and Beams in the USSR,"* **Rand Corporation, R-2816-ARPA, 1981.**

n **1985**, the **Budker Institute of Nuclear Physics (BINP)** signed a contract with **TsNIIMash** and the **Reshetnev Company** to develop **electron and ion sources** for simulating **the radiation effects of space environments on satellites**. The contract was worth **12 million Soviet rubles (MP)**.

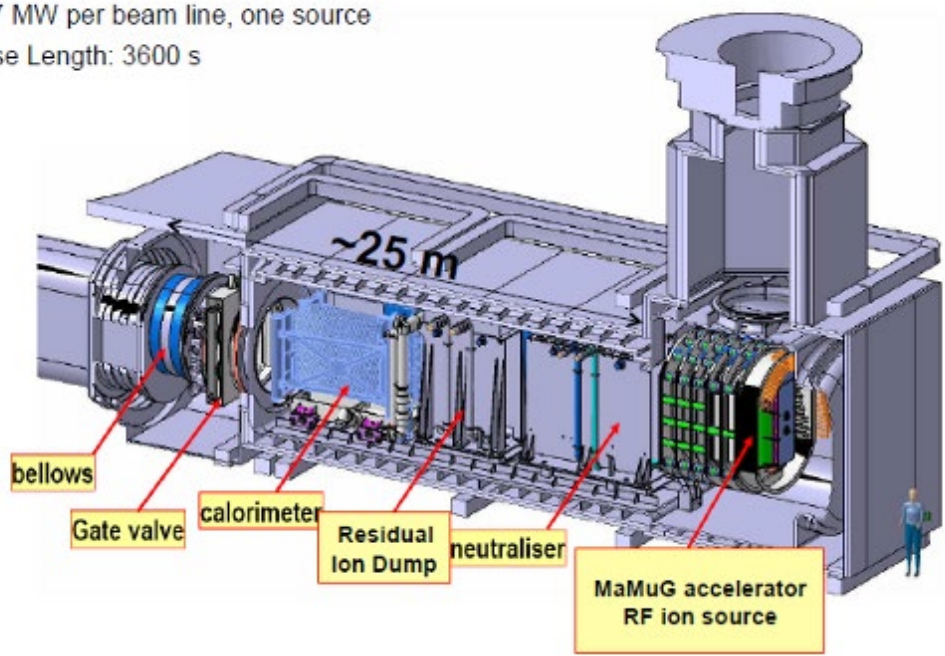

**Fig. 8. Structural Diagram of the Neutral Beam Injector with RF SPS for ITER**

The execution of this contract was assigned to **my laboratory**.

By **1990**, the contract was **successfully completed**, and the institute received **12 million full-value rubles**.

With this funding, **BINP** ordered **100 sets of industrial accelerators** from the **Ilyich Plant in Moscow**. These accelerators were later assembled into **commercial industrial accelerators** and **sold abroad for U.S. dollars**.

This financial move **helped the institute survive the difficult 1990s**.

The **12 million rubles** were **effectively converted into 100 million U.S. dollars**.

4o

At the Institute of Nuclear Physics (ИЯФ), many specialists in surface-plasma sources (SPS) have been trained, who are now working in various institutions such as IYAI, NIIEFA, and IAE.

The development of surface-plasma sources is described in the following books:

- **V. G. Dudnikov,** *Negative Ion Sources* , Novosibirsk State University, Novosibirsk, 2019.
- **Vadim Dudnikov,** *Development and Applications of Negative Ion Sources*, Springer, 2019.
- **Vadim Dudnikov,** *Development and Applications of Negative Ion Sources*, (second edition) Springer, 2023.
- V. Dudnikov, Surface plasma method of negative ion beam production, Phys. Uspekhi, 189, 12, 405, 2019.
- V. Dudnikov, Charge exchange injection in to accelerators and storage rings, Phys. Uspekhi, 189, 4, 433, 2019.
- 

..